\documentclass[aps,prl,reprint,showpacs,superscriptaddress]{revtex4-1}
\usepackage{graphicx}
\usepackage{color}
\usepackage{txfonts}
\usepackage[colorlinks, linkcolor=blue, anchorcolor=blue, citecolor=blue, urlcolor=blue]{hyperref}
\usepackage{multirow}

\begin{document}

% Use the \preprint command to place your local institutional report
% number in the upper righthand corner of the title page in preprint mode.
% Multiple \preprint commands are allowed.
% Use the 'preprintnumbers' class option to override journal defaults
% to display numbers if necessary
%\preprint{}

%Title of paper
\title{Distinct magnetic ground states of \textit{R}$_2$ZnIrO$_6$ (\textit{R} = La and Nd) determined by neutron powder diffraction}

% repeat the \author .. \affiliation  etc. as needed
% \email, \thanks, \homepage, \altaffiliation all apply to the current
% author. Explanatory text should go in the []'s, actual e-mail
% address or url should go in the {}'s for \email and \homepage.
% Please use the appropriate macro foreach each type of information

% \affiliation command applies to all authors since the last
% \affiliation command. The \affiliation command should follow the
% other information
% \affiliation can be followed by \email, \homepage, \thanks as well.

\author{H. Guo}
\email[]{hjguo@sslab.org.cn}
\affiliation{Neutron Science Platform, Songshan Lake Materials Laboratory, Dongguan, Guangdong 523808, China}

\author{C. Ritter}
\affiliation{Institut Laue-Langevin, Boite Postale 156X, F-38042 Grenoble Cedex 9, France}

\author{Y. Su}
\affiliation{J\"ulich Centre for Neutron Science (JCNS) at Heinz Maier-Leibnitz Zentrum (MLZ), Forschungszentrum J\"ulich, Lichtenbergstrasse 1, D-85747 Garching, Germany}

\author{A. C. Komarek}
\affiliation{Max-Planck-Institute for Chemical Physics of Solids, N\"{o}thnitzer Str. 40, D-01187 Dresden, Germany}

\author{J. S. Gardner}
\affiliation{Neutron Science Platform, Songshan Lake Materials Laboratory, Dongguan, Guangdong 523808, China}

%\author{}
%\email[]{Your e-mail address}
%\homepage[]{Your web page}
%\thanks{}
%\altaffiliation{}
%\affiliation{}

%Collaboration name if desired (requires use of superscriptaddress
%option in \documentclass). \noaffiliation is required (may also be
%used with the \author command).
%\collaboration can be followed by \email, \homepage, \thanks as well.
%\collaboration{}
%\noaffiliation

\date{\today}

\begin{abstract}
 Double perovskite iridates \textit{A}$_2$ZnIrO$_6$ (\textit{A} = alkaline or lanthanide) show complex magnetic behaviors ranging from weak ferromagnetism to successive antiferromagnetic transitions. Here we report the static (\textit{dc}) and dynamic (\textit{ac}) magnetic susceptibility, and neutron powder diffraction measurements for \textit{A} = La and Nd compounds to elucidate the magnetic ground state. Below 10~K, the \textit{A} = La compound is best described as canted iridium moments in an antiferromagnet arrangement with a propagation vector \textbf{k} = 0 and a net ferromagnetic component along the \textit{c}-axis. On the other hand, Nd$_2$ZnIrO$_6$ is described well as an antiferromagnet with a propagation vector \textbf{k} = (1/2~1/2~0) below $T_\mathrm{N} \sim$ 17 K. Scattering from both the Nd and Ir magnetic sublattices were required to describe the data and both were found to lie almost completely within the \textit{ab}-plane.   \textit{Dc} susceptibility revealed a bifurcation between the zero-field-cooled and field-cooled curves below $\sim$13 K in Nd$_2$ZnIrO$_6$. A glassy state was ruled out by \textit{ac} susceptibility but detailed magnetic isotherms revealed the opening of the loop below 13~K.  These results suggest a delicate balance exists between the Dzyaloshinskii-Moriya, crystal field schemes, and \textit{d}-\textit{f} interaction in this series of compounds.
\end{abstract}

% insert suggested PACS numbers in braces on next line
% insert suggested keywords - APS authors don't need to do this
%\keywords{}

%\maketitle must follow title, authors, abstract, \pacs, and \keywords
\maketitle

% body of paper here - Use proper section commands
% References should be done using the \cite, \ref, and \label commands

Double perovskite with the general formula \textit{A}$_2$\textit{B'B''}O$_6$ forms an ordered rock-salt-like structure when the \textit{B'} and \textit{B''} ions are physically and/or chemically significantly different. In general, the \textit{A} site is occupied by an alkaline or lanthanide cation and \textit{B'}/\textit{B''} are transition-metal elements. This structure is incredibly versatile, hosting a wide variety of ions on the \textit{A}, \textit{B'}, and \textit{B''} sites. The interplay between crystal structure, orbital and spin degrees of freedom, and electron-lattice coupling leads to a diverse range of physical properties from metal-insulator transition \cite{Kato-2002} to multiferroicity \cite{Shimakawa-2011}, and large magnetoresistance effect \cite{Kobayashi-1998} to photolysis \cite{Luo-2014}. For a single magnetic \textit{B'}/\textit{B''}-site compound, the superexchange coupling between two nearest cations through intermediate oxygen takes part in the magnetic order. Generally, when the \textit{A} site is occupied by a magnetic cation, the trivalent lanthanide orders at a lower temperature than the transition-metal sublattice \cite{Ding-2019,Sanchez-2011}.
Recently, double perovskites with Ir ions at the \textit{B''} site have attracted much attention \cite{Cao-2013,Zhu-2015,Aczel-2016,Vogl-2018,Ding-2019,Singh-2020,Nd,Gao-2020} due to the relatively large spin-orbit coupling (SOC) of the Ir$^{4+}$ ion which may lead to novel phenomena such as SOC-driven Mott insulator \cite{Kim-2008} and complex magnetism \cite{Guo-2013,Guo_2016,Guo-2017}.
Some \textit{A}$_2$\textit{B'}IrO$_6$ have been synthesized but the magnetic ground states are still not well understood. La$_2$MgIrO$_6$ shows antiferromagnetic behavior below $T_\mathrm{N}$ = 12 K \cite{Cao-2013}. Isovalent substitution of Zn for Mg results in a different magnetic ground state; where the magnetic susceptibility increases substantially below 10 K and hysteresis is observed in magnetic isothermals below this temperature, suggesting a ferromagnetic component to the magnetic structure \cite{Cao-2013}.
More recent studies on small single crystals suggested a canted antiferromagnetic state with a net moment along the crystallographic \textit{b}-axis \cite{Han-2018}. Introducing rare earth moments on the \textit{A} site gives rise to complex magnetic properties \cite{Nd}, and in Nd$_2$ZnIrO$_6$ two transitions have been observed in the magnetic susceptibility below $\sim$17 K. Preliminary neutron powder diffraction (NPD) measurements exist for these compounds, but a reliable model to explain the materials with two magnetic sublattices has so far eluded scientists. The relatively large neutron absorption by Ir ions and their comparatively small magnetic moment limits the statistics and reduces the uniqueness of models.

In this paper, we have performed NPD measurements, as well as bulk \textit{dc} and \textit{ac} magnetic susceptibility measurements on polycrystalline La$_2$ZnIrO$_6$ and Nd$_2$ZnIrO$_6$ in order to investigate the static magnetic ground state. La$_2$ZnIrO$_6$ forms a canted antiferromagnetic structure where the moments are constrained within the \textit{bc}-plane with a finite ferromagnetic component along the long \textit{c}-axis. This result differs from the earlier neutron diffraction and first principle calculations that did not detect a ferromagnetic component \cite{Cao-2013}. Our results are similar to, but also inconsistent with, the single crystal magnetisation data that found the easy axis along the crystallographic \textit{b}-axis \cite{Han-2018}. Substituting the nonmagnetic La$^{3+}$ with the magnetic Nd$^{3+}$, we found long range antiferromagnetic order below 17 K in Nd$_2$ZnIrO$_6$. Data at 1.8~K is described well as an antiferromaget with the moments (Nd and Ir) lying almost exactly within the \textit{ab}-plane and with no net moment. Interestingly, only one transition was detected in both NPD and \textit{ac} susceptibility measurements at $T_\mathrm{N} \sim$ 17 K, while two are seen in the bulk static susceptibility. The distinct ground states of these two double perovskites reveal the delicate balance between the Dzyaloshinskii-Mariya interaction and \textit{d}-\textit{f} superexchange interaction within these compounds.

Polycrystalline samples of La$_2$ZnIrO$_6$ and Nd$_2$ZnIrO$_6$ were prepared by solid state reaction method. Stoichiometric amounts of raw materials of La$_2$O$_3$, Nd$_2$O$_3$, ZnO and IrO$_2$ were mixed and ground thoroughly in an agate mortar. The powders were then pressed into pellets and sintered between 900$^{\circ}$C and 1100$^{\circ}$C in air for about 20 days with several intermediate grindings. Phase purity was checked by x-ray diffraction (XRD) and NPD measurements. Magnetization measurements were performed using the vibrating sample magnetometer (MPMS3, Quantum Design). \textit{AC} susceptibility was measured in the Physical Property Measurement System (PPMS, Quantum Design) with an \textit{ac} excitation field of 5 Oe.
NPD measurements were performed at the Institut Laue-Langevin (ILL) in Grenoble, France. High Q-resolution measurements, for crystal structure determination, were performed on the D2B diffractometer with 1.594 \AA~ neutrons selected by a Ge(335) monochromator, while the D20 diffractometer produced a higher flux beam of 2.394 \AA~ neutrons, albeit with lower resolution suitable for magnetic structure refinements. Approximately 7-g of samples were loaded into annular vanadium cans, to reduce the neutron flight path within the sample, overcoming the high neutron absorption from naturally abundant Ir. The La$_2$ZnIrO$_6$ sample was measured on the D20 diffractometer for 10.25 h at 20 K and 1.8 K. Due to the stronger magnetic signal from the Nd$_2$ZnIrO$_6$ sample, 2-hour measurements at (1.8, 13.5, 15.1 and 19.3) K, were performed.
In addition, a series of 15-min measurements were carried out while ramping the temperature at 0.1 K/200sec and 0.1 K/450sec between (1.8 - 12 )K and (12 - 20)K respectively. These data allowed the temperature dependence of the magnetic peak intensities to be determined, which is proportional to the magnetization squared. To determine the magnetic contribution to the NPD, a high temperature (paramagnetic) data set was subtracted from that collected at lower temperatures. Rietveld refinements were performed using the FullProf software suite \cite{Fullprof} together with SARAh for symmetry analysis \cite{sarah}.

\begin{figure}
\centering
\includegraphics[width=1\columnwidth]{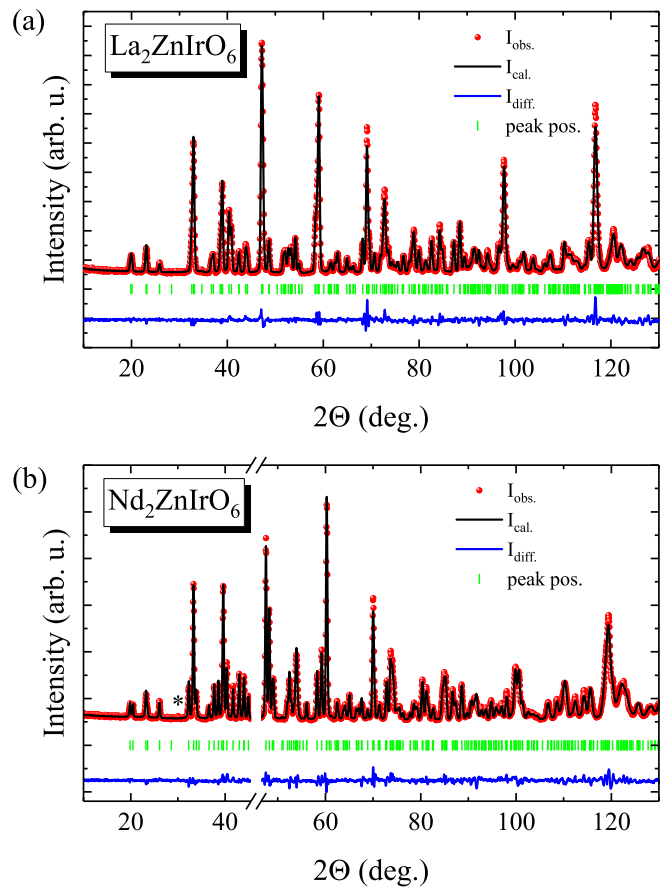}
\caption{(Color online) Rietveld refined patterns from $R_2$ZnIrO$_6$. This high resolution data was collected at room temperature on the D2B diffractometer. The lattice parameters are \textit{a} = 5.5906(2) \AA, \textit{b} = 5.6871(2) \AA, \textit{c} = 7.9312(2) \AA, $\beta$ = 90.00(1)$^{\circ}$ for \textit{R} = La sample and \textit{a} = 5.4697(2) \AA, \textit{b} = 5.7313(2) \AA, \textit{c} = 7.8058(2) \AA, $\beta$ = 89.97(1)$^{\circ}$ for the \textit{R} = Nd sample. A small peak originated from the cryofurnace has been excluded from the refinement in (b), and the asterisk marks the peak originated from the Nd$_2$Ir$_2$O$_7$ phase.}
\label{nuc_refinement}
\end{figure}

Both XRD and NPD measurements indicate an impurity-free phase for the La$_2$ZnIrO$_6$ sample, but a small amount of Nd$_2$Ir$_2$O$_7$ (about 0.1\textit{wt}\%) was found in the Nd$_2$ZnIrO$_6$ sample. Refinement of the crystal structure confirms both samples crystallize into a monoclinic structure with space group $P2_1/n$ (\#14); see Fig. \ref{nuc_refinement}. In addition, NPD measurements were able to investigate the presence of \textit{B}-site disorder. In La$_2$ZnIrO$_6$, this amounts to between 6(4)\% from the D2B measurement and 13(4)\% from the D20 measurement, as will be discussed later. Similar antisite disorder was found to be negligible in Nd$_2$ZnIrO$_6$.

\begin{figure}
\centering
\includegraphics[width=1\columnwidth]{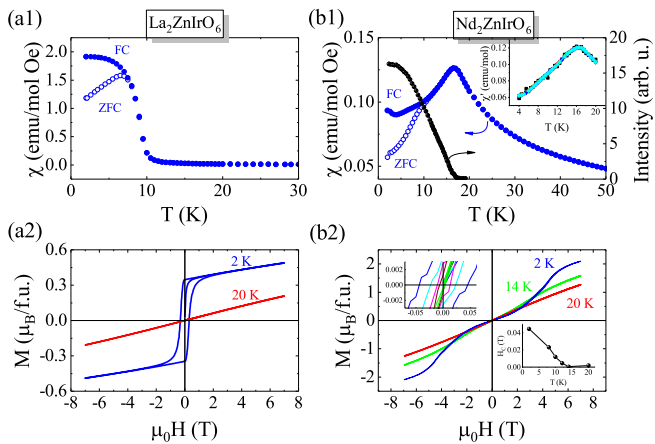}
\caption{(Color online) Temperature dependence of the magnetic susceptibility and isothermal magnetization measurements for (a1, a2) La$_2$ZnIrO$_6$ and (b1, b2) Nd$_2$ZnIrO$_6$. The susceptibility was measured with a field of 1000 Oe. The temperature dependence of the magnetic peak intensity, measured on D20, is also shown in (b1) where the 13~K transition is absent. The inset of (b1) shows the real component ($\chi'$) of the zero field \textit{ac} susceptibility with frequencies of 13, 113, 1333, 5330 and 9918 Hz. The left inset of (b2) is an enlargement of the low field region measured at 2 (blue), 8 (cyan), 10 (magenta), 12 (wine) and 14 (green) K. The right inset shows the temperature dependence of the coercive field ($H_c)$.}
\label{MT-MH}
\end{figure}

The temperature dependence of the magnetic susceptibility and isothermal magnetization measurements are shown in Fig. \ref{MT-MH}. For La$_2$ZnIrO$_6$, the magnetic susceptibility increases abruptly below 10 K, concomitant with a splitting between the zero-field-cooled (ZFC) and field-cooled (FC) magnetisation curves and hysteresis in the isothermal magnetization loop, suggesting a weak ferromagnetic component in the ground state.
On the other hand, the Nd$_2$ZnIrO$_6$ sample shows a more complicated path to the ground state. A broad peak can be found at 17 K below which long range antiferromagnetic order is observed, but a second magnetic phase is revealed in the splitting between the ZFC and FC magnetization curves below 13 K. Note that the magnitude of the susceptibility in Nd$_2$ZnIrO$_6$ is much smaller than that of the La$_2$ZnIrO$_6$ sample, although the net magnetic moment is larger (see Fig. \ref{MT-MH} (a2) and (b2)) due to the much larger Nd$^{3+}$ moment. These bulk results are consistent with those reported in the literature \cite{Cao-2013,Nd}.
Temperature dependence of the real component of the dynamic susceptibility of Nd$_2$ZnIrO$_6$ is shown in the inset of Fig. \ref{MT-MH}(b1) when measured in a zero DC field.  These data reveal no dependence on excitation frequency excluding a possible spin-glass like origin to the 13~K anomaly in the static magnetization similar to that seen in the double perovskite La$_2$CoIrO$_6$ by Song {\it et al.} \cite{Song-2017}.  In order to study the magnetic phase below 13~K further, careful magnetic isotherms were collected around this temperature up to 7 T.  Unlike  La$_2$ZnIrO$_6$ (Fig. \ref{MT-MH}(a2)) where an antiferromagnet develops a ferromagnetic component and a loop opens up around the origin below the critical temperature, in Nd$_2$ZnIrO$_6$ below the high temperature transition no gap is seen in the magnetisation loops, consistent with a simple antiferromagnet.  However, below the second transition at 13~K, the loops take on a "S" shape, typical of a metamagnet and on closer inspection ( upper inset Fig. \ref{MT-MH}(b2)) a small gap opens up around the origin, suggestive of a ferromagnetic component to the ordering, similar to that in La$_2$ZnIrO$_6$.

\begin{figure}
\centering
\includegraphics[width=1\columnwidth]{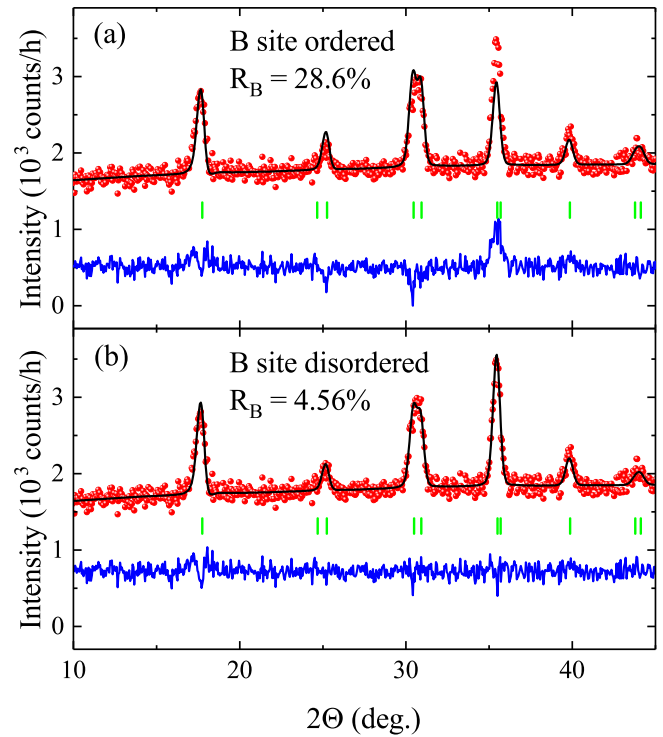}
\caption{(Color online) Rietveld refinement of the magnetic diffraction from La$_2$ZnIrO$_6$ at 1.8 K. A paramagnetic data set, taken at 20 K has been subtracted to remove the scattering from the crystal lattice. The calculated patterns are according to the IR $\Gamma_3$ with (a) a perfectly ordered \textit{B}-site model and (b) a \textit{B}-site disordered model where the best model found 13(4)\% site substitution between Zn and Ir ions. Red dot: experimental data; Black curve: calculated data; Vertical bars: magnetic peak position; Blue curve: difference between the experimental and calculated data.}
\label{La_refine}
\end{figure}

\begin{figure}
\centering
\includegraphics[width=1\columnwidth]{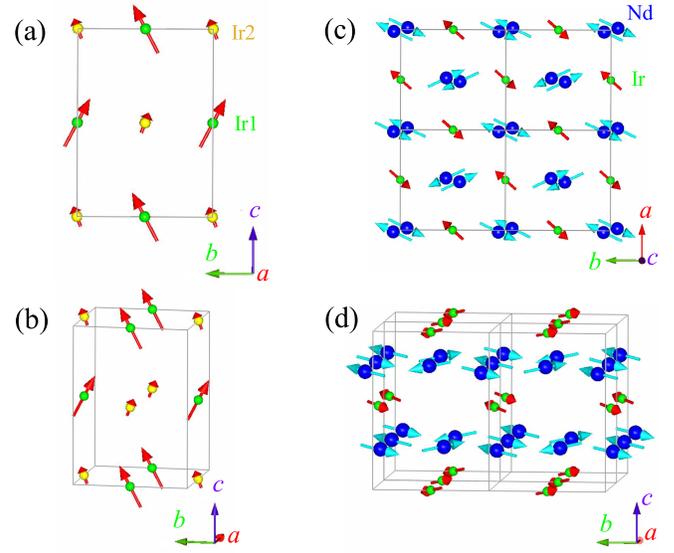}
\caption{(Color online) Magnetic structure of (a, b) La$_2$ZnIrO$_6$ and (c, d) Nd$_2$ZnIrO$_6$ viewed along different directions. The Ir moments have been enlarged for clarity. Note that the smaller Ir2 moments in (a, b) is visualized taking into account the partial occupation effect. }
\label{magstr}
\end{figure}

In order to determine the underlying magnetic structure for both compounds, we have performed NPD measurements on the high flux D20 diffractometer at the ILL in the absence of a magnetic field. Atomic structure refinements were performed at 20 K in the paramagnetic state, allowing the scale factors to be determined and fixed for subsequent magnetic structure refinements. All measurements were performed after cooling in zero field. We first concentrate on the La$_2$ZnIrO$_6$ sample, where the magnetic peaks can be indexed with a \textbf{k} = 0 propagation vector using the K-search program. These data and the magnetic structure refinements confirmed the Zn/Ir mixing (\textit{B}-site disorder) previously observed in the crystallographic data collected on D2B. Magnetic symmetry analysis for the Ir1 moment (2\textit{c} site) and the cation disordered Ir2 moment on the Zn position (2\textit{d} site) with the monoclinic space group $P2_1/n$ shows that the reducible magnetic representation is decomposed into two IRs as: $\Gamma = \Gamma_1 + \Gamma_3$. The basis vectors (BV) for each IR are shown in Supplementary Materials (SM) Tab. S1. The magnetic intensities can be well described by IR $\Gamma_3$.
As can be seen from Fig. \ref{La_refine}, a \textit{B}-site ordered model results in a poorer fit and $R_B$ value compared to the \textit{B}-site disordered model. Note that in order to have the same magnitude of magnetic moment at the Ir1 and Ir2 atoms, the degree of disorder obtained from the magnetic refinement is $\sim$~13(4)\%,  which is slightly larger than the 6(4)\% found from the D2B refinement.
The obtained magnetic structure is shown in Fig. \ref{magstr}(a,b). The magnitude of the moment is 0.87 $\mu_\mathrm{B}$/Ir at 1.8 K and they are lying predominately within the \textit{bc}-plane with $\mu_b$ = 0.403(7) $\mu_\mathrm{B}$ and $\mu_c$ = 0.773(8) $\mu_\mathrm{B}$ at \textit{B''} and $\mu_b$ = 0.30(5) $\mu_\mathrm{B}$ and $\mu_c$ = 0.82(5) $\mu_\mathrm{B}$ for the Ir that has accidentally substituted for Zn on the \textit{B'} site. As can be seen, there is a canting of these spins resulting in a net moment along the \textit{c}-axis, which explains the ferromagnetic component observed by bulk magnetic susceptibility measurements, see Fig. \ref{MT-MH}(a2) at base temperature.
Single crystal magnetization measurements led Han \textit{et al}. to conclude that the \textit{b}-axis was the easy axis \cite{Han-2018}, which would correspond to the IR $\Gamma_1$ (see SM Tab. S1). However, the IR $\Gamma_1$ does not describe our neutron data, see Fig. S1 in the SM. Also incompatible with our data, Cao \textit{et al}. suggested that the moments lie within the \textit{ab}-plane by LDA + U calculations \cite{Cao-2013}, which need to be revisited in light of our conclusions.

As mention above, the static magnetic susceptibility suggests a different magnetic ground state for Nd$_2$ZnIrO$_6$ compared to La$_2$ZnIrO$_6$. Additional magnetic peaks can be observed below $T_\mathrm{N} \sim$ 17 K, which can be indexed with a propagation vector \textbf{k} = (1/2~1/2~0), consistent with an earlier report \cite{Nd}. The temperature dependence of the integrated peak intensity at 2$\Theta \sim$ 25.1$^\circ$ monotonically increases as the temperature is lowered through $T_\mathrm{N}$ and no anomaly is seen at 13~K, where bulk magnetic susceptibility reveals a splitting of the ZFC and FC curves, see Fig. \ref{MT-MH}(b1).

\begin{figure}
\centering
\includegraphics[width=1\columnwidth]{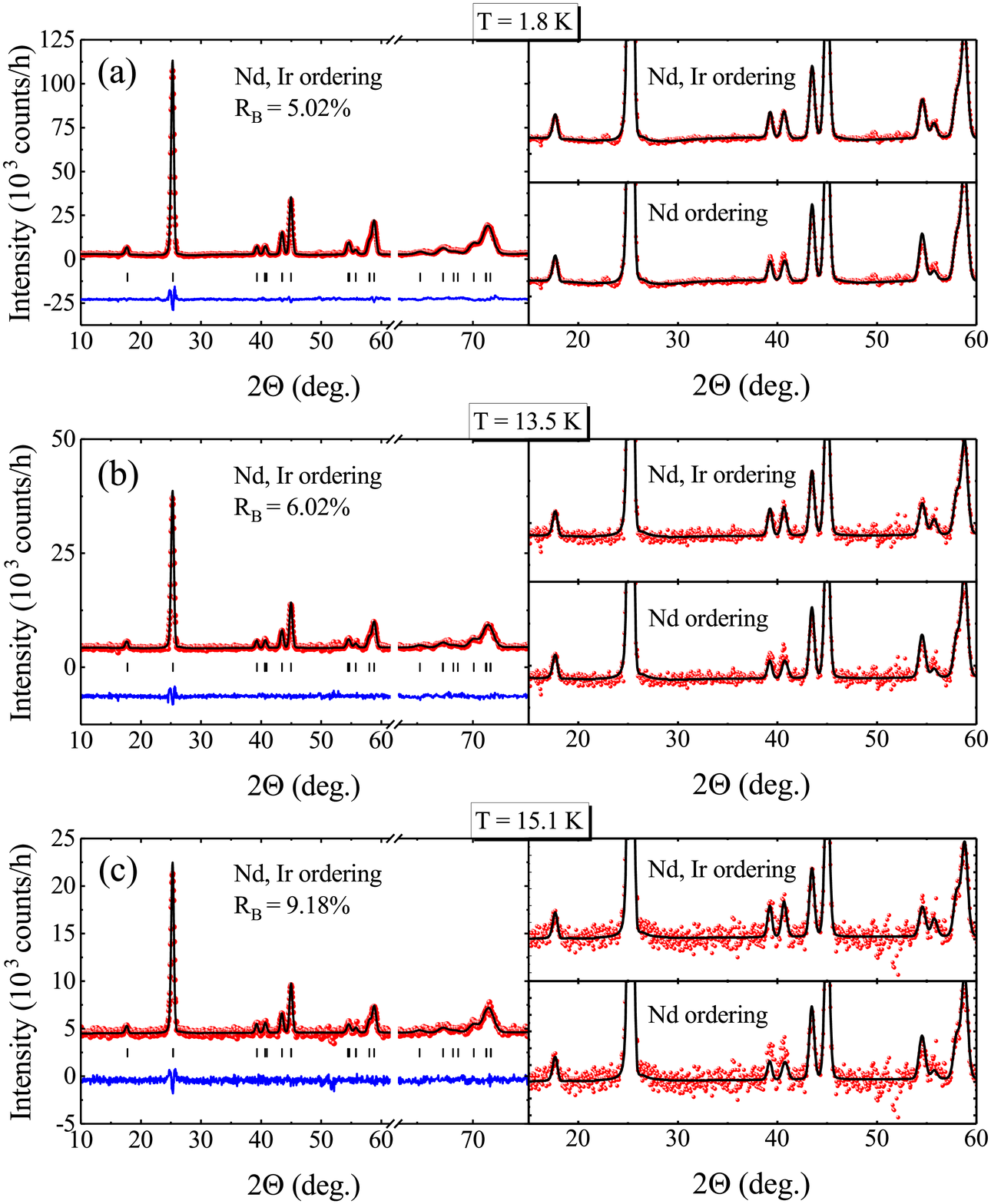}
\caption{(Color online) Rietveld magnetic refinement of the difference pattern at various temperatures for Nd$_2$ZnIrO$_6$ according to IR $\Gamma_1~+~\Gamma_3$. The left panels show the refinement results with perfectly ordered Nd and Ir sublattices, and the right panels highlights an expanded region of low peak intensities comparing our best model and that with only the Nd sublattice ordering. Red dot: experimental data; Black curve: calculated data; Vertical bars: magnetic peak position; Blue curve: difference between the experimental and calculated data. A small angular range around $2\Theta$ = 62$^{\circ}$ was excluded from the fits due to contamination from the atomic scattering that did not subtract well.}
\label{Nd_refine}
\end{figure}

Symmetry analysis for the Nd ions at the 4\textit{e} site and Ir ions at the 2\textit{c} site with the propagation vector k = (1/2 1/2 0) show that the reducible magnetic representations decompose as follows:
\begin{equation}\label{}
 \Gamma_\mathrm{Ir} = \Gamma_1 + \Gamma_3
\end{equation}
\begin{equation}\label{}
 \Gamma_\mathrm{Nd} = \Gamma_1 + \Gamma_2 + \Gamma_3 + \Gamma_4
\end{equation}

The corresponding BVs are listed in SM Tab. S2 and S3. As pointed out by an earlier report \cite{Nd}, half of the magnetic moments are not ordered for either of these IRs, and a combination of the two IRs is required in order to form a fully ordered state. In addition, if there exist a coupling between the two magnetic sublattices and the magnetic transition is second order in character \cite{Bertaut-1968} both sublattices should order under the same IR. An earlier report \cite{Nd} determined the ordering of the Nd$^{3+}$ moments, but Vogl $\it{et~al.}$ was not able to resolve the ordering of the Ir$^{4+}$ moments due to insufficient statistics. Using the high flux diffractometer, D20, at the ILL we are able to solve the magnetic structure of this compound which must include static moments on both the Nd and Ir sublattices to describe the data well. Consistent with the high resolution neutron diffraction data, no antisite disorder was required to describe these data.

As can be seen from SM Tab. S2, only one solution, i.e., $\Gamma_1 + \Gamma_3$ is possible. The best refinements with Nd and Ir sublattices ordering under $\Gamma_1 + \Gamma_3$ are shown in Fig. \ref{Nd_refine}. Focussing on the low intensity peaks in the right panels of Fig. \ref{Nd_refine}, it is clear that a model with only the Nd sublattice ordering cannot account for the peak intensities around $2\Theta \sim$ 40$^{\circ}$. This is more apparent close to $T_\mathrm{N}$ where the size of both the Nd and Ir moments are expected to be small and comparable, producing a more significant interference effect, as observed in the pyrochlore iridate Nd$_2$Ir$_2$O$_7$ \cite{Guo_2016}.
The refined magnetic moments at 1.8~K are listed in Tab. \ref{moment}, and the magnetic structure is shown in Fig. \ref{magstr}(c,d). The moments for both Nd and Ir ions prefer to lie within the \textit{ab}-plane, in contract to the La$_2$ZnIrO$_6$ case.

The value of the Nd moment is much smaller than the free ion value ($g_J$ = 8/11, \textit{J} = 9/2) presumably due to the crystal electric field effect as observed in similar compounds \cite{Aczel-2013}. It can be seen in Tab. \ref{moment} that although both magnetic sublattices develop concomitantly, the Nd moment at 15.1~K (2~K below the transition temperature) is only 39\% of that at 1.8~K, while the Ir moment is already 72\% of the low temperature value by 15.1~K. This observation suggests the Ir sublattice is the driving force of the magnetic ordering. The obtained antiferromagnetic structure has zero net moment along any crystallographic directions, thus, the origin of the weak ferromagnetic component seen in static magnetic susceptibility and FC / ZFC cooled splitting measurements is still a puzzle. This bulk susceptibility character seems to be ubiquitous in the iridates. The pyrochlore iridates show all-in/all-out magnetic structure with zero net moment, whereas the static magnetic susceptibility also show the bifurcation below $T_\mathrm{N}$, and was attributed to the formation of antiferromagnetic domain walls \cite{Matsuhira-2011,Ma-2015,Hirose-2017}. One difference is that the bifurcation occurs at $T_\mathrm{N}$ for the pyrochlore iridate, while it appears at lower temperature for Nd$_2$ZnIrO$_6$. Another possibility is a subtle change of the magnetic symmetry at this temperature which could result in a small net moment, this was not detected in our current study, but may require higher resolution, single crystal magnetic structural refinement studies. The former scenario would correspond to a phase separation, while the latter would occur within one phase. Future studies using local probe would be helpful to elucidate this point.

To conclude we have determined with the aid of neutron powder diffraction and symmetry analysis the long range magnetic structure of La$_2$ZnIrO$_6$, albeit with approximately $\frac{1}{8}$ disorder on the \textit{B}-site. This simple \textbf{k} = 0 structure has a net moment along the \textit{c}-axis. Further studies are needed to elucidate the role of site disorder in these findings. Adding Nd$^{3+}$ with a magnetic moment to the rare-earth site results in a significant change to the chemistry and bulk magnetic properties. No crystallographic disorder was found in this sample, and all moments are predominantly in the \textit{ab}-plane. The iridium sublattice was found to drive the magnetic transition, but the 5\textit{d}-4\textit{f} interactions are clearly important, increasing the ordering temperature by 70\%. Moreover, the single ion anisotropy of the Nd$^{3+}$ ion probably governing the spin direction, in this case rotating the preferred plane from the \textit{bc}-plane to the \textit{ab}-plane.

Double perovskites have been shown to be an excellent play ground to investigate the interplay of spin-lattice and electron degrees of freedom. With the wide array of chemical diversity many double perovskites have also been shown to be technologically relevant. We hope this study motivates others to study Ir$^{4+}$ based double perovskites where spin-orbit coupling, on-site Coulomb interaction, and crystal field energies are of comparable energy. For example in Nd$_2$ZnIrO$_6$ the isothermal magnetization curve at 2~K has the signature of a metamagnetic transition at about 2 T which should be studied in more detail, but this is true for many of the iridate based double perovskites\cite{Ding-2019}.

\begin{table}
\caption{Moment size for the Nd and Ir ions at different temperatures according to the best refinement assuming both sublattices order under the same IR $\Gamma_1 + \Gamma_3$. The unit of the moment is $\mu_\mathrm{B}.$\label{moment}}
\begin{ruledtabular}
\begin{tabular}{lcccc}
\multicolumn{2}{c}{}& 1.8 K & 13.5 K & 15.1 K \\
\hline
\multirow{3}{*}{Nd} & $\mu_a$ & 0.94(2) & 0.47(2) & 0.27(3)  \\
           & $\mu_b$ & 1.98(1) & 1.10(1) & 0.78(1)  \\
           & $\mu_c$ & 0.38(7) & 0.24(4) & 0.24(4)  \\
           & $|\mu|$ & 2.23(2) & 1.22(1) & 0.87(2) \\
\hline
\multirow{3}{*}{Ir} & $\mu_a$ & -0.47(5) & -0.41(4) & -0.37(5) \\
           & $\mu_b$ & -0.53(4) & -0.40(3) & -0.36(3) \\
           & $\mu_c$ & 0.15(4) & 0.10(3) & 0.08(4)  \\
           & $|\mu|$ & 0.72(4) & 0.58(4) & 0.52(4) \\
\end{tabular}
\end{ruledtabular}
\end{table}

\bibliography{R2ZnIrO6}

\section{Supplemental Material}
\renewcommand{\thetable}{S\arabic{table}}
\renewcommand{\thefigure}{S\arabic{figure}}
\setcounter{figure}{0} 
\setcounter{table}{0}

\begin{figure}[b]
\centering
\includegraphics[width=0.8\columnwidth]{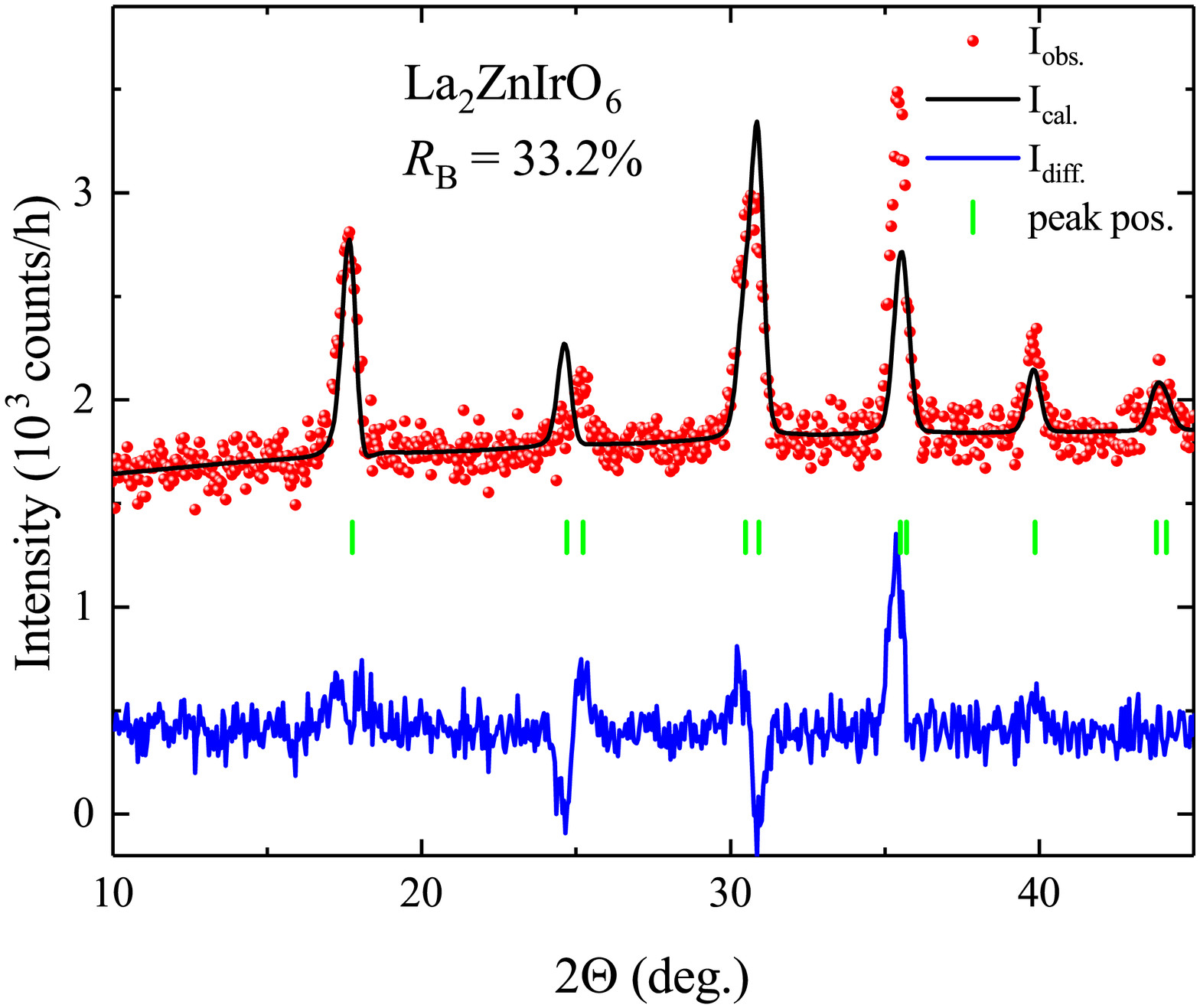}
\caption{(Color online) Magnetic structure refinement for La$_2$ZnIrO$_6$ according to IR $\Gamma_1$.}
\label{s1}
\end{figure}

Magnetic structure refinement for the La$_2$ZnIrO$_6$ compound according to IR $\Gamma_1$ is shown in Fig. \ref{s1}, which would have a nonzero ferromagnetic component along the \textit{b}-axis. As can be seen, this model cannot describe the overall peak intensities satisfactorily, thus, resulting in a large agreement factor $R_B$ = 33.2\%. Even when disorder is included in the model, the value of $R_B$ is still as large as 22.0\%. Moreover, the degree of \textit{B}-site disorder amounts to 25\%, which is in contradiction to the D2B measurements.

\begin{table}[b]
\caption{Irreducible representations (IR) and the basis vectors $\varphi$ for the Ir ions at the 2\textit{c} and 2\textit{d} sites with the  space group $P2_1/n$ (\#14) and propagation vector \textbf{k}~=~(0 0 0). Site1: (x, y, z); Site2: (-x+1/2, y-1/2, -z+1/2).\label{BV1}}
\begin{ruledtabular}
\begin{tabular}{lccc}
IRs & $\varphi$  & Site1 & Site2 \\
\hline
$\Gamma_1$ & $\varphi_1$  & (1~0~0) & (-1~0~0)  \\
           & $\varphi_2$  & (0~1~0) & (0~1~0)  \\
           & $\varphi_3$  & (0~0~1) & (0~0~-1)  \\
$\Gamma_3$ & $\varphi_4$  & (1~0~0) & (1~0~0) \\
           & $\varphi_5$  & (0~1~0) & (0~-1~0) \\
           & $\varphi_6$  & (0~0~1) & (0~0~1) \\
\end{tabular}
\end{ruledtabular}
\end{table}

\begin{table}
\caption{Irreducible representations (IR) and the basis vectors $\varphi$ for the Ir ions at the 2\textit{c} site with space group $P2_1/n$ (\#14) and propagation vector \textbf{k}~=~(1/2 1/2 0). Site1: (x, y, z); Site2: (-x+1/2, y-1/2, -z+1/2).\label{BV2}}
\begin{ruledtabular}
\begin{tabular}{lccc}
IRs & $\varphi$  & Site1 & Site2  \\
\hline
$\Gamma_1$ & $\varphi_1$  & (1~0~0) & (0~0~0)  \\
           & $\varphi_2$  & (0~1~0) & (0~0~0)  \\
           & $\varphi_3$  & (0~0~1) & (0~0~0)  \\
$\Gamma_3$ & $\varphi_4$  & (0~0~0) & (-1~0~0) \\
           & $\varphi_5$  & (0~0~0) & (0~1~0) \\
           & $\varphi_6$  & (0~0~0) & (0~0~-1) \\
\end{tabular}
\end{ruledtabular}
\end{table}

\begin{table}
\caption{Irreducible representations (IR) and the basis vectors $\varphi$ for the Nd ions at the 4\textit{e} site with space group $P2_1/n$ (\#14) and propagation vector \textbf{k}~=~(1/2 1/2 0). Site1: (x, y, z); Site2: (-x+1/2, y-1/2, -z+1/2); Site3: (-x+1, -y+1, -z+1); Site4: (x+1/2, -y+3/2, z+1/2). \label{BV3}}
\begin{ruledtabular}
\begin{tabular}{lccccc}
IRs & $\varphi$  & Site1 & Site2 & Site3 & Site4 \\
\hline
$\Gamma_1$ & $\varphi_1$  & (1~0~0) & (0~0~0)  & (-1~0~0) & (0~0~0)  \\
           & $\varphi_2$  & (0~1~0) & (0~0~0)  & (0~-1~0) & (0~0~0)  \\
           & $\varphi_3$  & (0~0~1) & (0~0~0)  & (0~0~-1) & (0~0~0)  \\
$\Gamma_2$ & $\varphi_4$  & (1~0~0) & (0~0~0)  & (1~0~0) & (0~0~0)  \\
           & $\varphi_5$  & (0~1~0) & (0~0~0)  & (0~1~0) & (0~0~0)  \\
           & $\varphi_6$  & (0~0~1) & (0~0~0)  & (0~0~1) & (0~0~0)  \\
$\Gamma_3$ & $\varphi_7$  & (0~0~0) & (-1~0~0)  & (0~0~0) & (1~0~0)  \\
           & $\varphi_8$  & (0~0~0) & (0~1~0)  & (0~0~0) & (0~-1~0)  \\
           & $\varphi_9$  & (0~0~0) & (0~0~-1)  & (0~0~0) & (0~0~1)  \\
$\Gamma_4$ & $\varphi_{10}$  & (0~0~0) & (-1~0~0)  & (0~0~0) & (-1~0~0)  \\
           & $\varphi_{11}$  & (0~0~0) & (0~1~0)  & (0~0~0) & (0~1~0)  \\
           & $\varphi_{12}$  & (0~0~0) & (0~0~-1)  & (0~0~0) & (0~0~-1)  \\
\end{tabular}
\end{ruledtabular}
\end{table}

\begin{acknowledgments}

\end{acknowledgments}

\end{document}